\documentclass[prb,preprint]{revtex4-1} 
\usepackage{amsmath}
\usepackage{lipsum}
\usepackage{graphicx}
\usepackage{epstopdf}
\usepackage{dcolumn}
\usepackage{bm}
\usepackage{ulem} 
\usepackage[usenames]{color}

\usepackage{epsfig}
\usepackage{float}
\usepackage{subfigure}
\usepackage{cancel}
\usepackage{multirow}
\usepackage[version=3]{mhchem}

\usepackage{soul}

\newcommand{\nc}{\newcommand}       
\nc{\vc}[1] {\mbox{\boldmath $#1$}} 
\nc{\del}       {\partial}              
\nc{\bra}       {\langle}               
\nc{\ket}       {\rangle}               
\nc{\bras}[1]   {\langle #1|}           
\nc{\kets}[1]   {|#1\rangle}            
\nc{\mapleft}[1]{           
	\smash{\mathop{\,          %
			\hbox to 1.5cm{\rightarrowfill}\, }\limits_{#1}}}
\nc{\beq}     {\begin{eqnarray}} \nc{\eeq}    {\end{eqnarray}}
\nc{\nn}      {\\\nonumber} \nc{\vs}      {\vspace{-0.275cm}}
\nc{\fra}    {\frac{1}{2}}
\nc{\mb}        {\mathbf}

\begin{document}
	
	\preprint{}
	
	\title{One-dimension Periodic Potentials in Schr\"odinger Equation Solved by the Finite Difference Method}
	\author{Lingfeng Li}
	\affiliation{School of Physics, Nankai University, Tianjin 300071,  China}
	
	\author{Jinniu Hu}~\email{hujinniu@nankai.edu.cn}
	\affiliation{School of Physics, Nankai University, Tianjin 300071,  China}
	\affiliation{Shenzhen Research Institute of Nankai University, Shenzhen 518083, China}
	
	\author{Ying Zhang}
	\affiliation{Department of Physics, School of Science, Tianjin University, Tianjin 300354, China}

\date{\today}
\begin{abstract}
The one-dimensional Kronig-Penney potential in the Schr\"{o}dinger equation, a standard periodic potential in quantum mechanics textbooks known for generating band structures, is solved by using the finite difference method with periodic boundary conditions. This method significantly improves the eigenvalue accuracy compared to existing approaches such as the filter method. The effects of the width and height of the Kronig-Penney potential on the eigenvalues and wave functions are then analyzed. As the potential height increases, the variation of eigenvalues with the wave vector slows down. Additionally, for higher-order band structures, the magnitude of the eigenvalue significantly decreases with increasing potential width. Finally, the Dirac comb potential, a periodic $\delta$ potential, is examined using the present framework. This potential corresponds to the Kronig-Penney potential's width and height approaching zero and infinity, respectively. The numerical results obtained by the finite difference method for the Dirac comb potential are also perfectly consistent with the analytical solution.
\end{abstract}



\maketitle

\section{Introduction}

The Schr\"{o}dinger equation is a frequently used equation to describe the motion of microscopic particles. Usually in the time-independent Schr\"{o}dinger equation, with a given potential, people wants to know the eigenvalues of the energy and the corresponding eigen wave functions.  However, there are only a few potentials that can be analytically solved in the Schr\"{o}dinger equation, such as the infinite well potential, harmonic oscillator potential, Coulomb potential, and so on, as shown in the available textbooks of quantum mechanics~\cite{zettili2009,greiner2011,shankar2012,griffiths2018}. Many commonly used potentials in physics must be solved with numerical methods. With the development of computational physics, many numerical schemes have been proposed to solve the Schr\"{o}dinger equation.

The shooting method is the most commonly used approach to solve the eigenvalue problem with boundary conditions \cite{horowitz1981,killingbeck1987}. The imaginary time step method can generate the real wave function with an imaginary time evolution factor \cite{goldberg1967,zhang2010}. The Green's function method can obtain the bound and resonant states as poles in the complex energy plane \cite{zhang2011}. In the basis expansion method \cite{shore1973,gambhir1990}, the exact wave function is expanded by an orthogonal complete set. The expansion coefficients and eigenvalues can be obtained by diagonalizing the matrix, which is constructed by the expectation values of the Hamiltonian between different basis wave functions. The finite element method divides the original solution domain into a finite number of subdomains, which are called elements and selects polynomial functions as the basis~\cite{levin1985,fischer1993}. The finite difference method (FDM) can directly express the second order derivative in the kinetic energy operator of the Schr\"{o}dinger equation as a discretized form in the coordinate space \cite{simos1997}, which we recently successfully applied to solve the spurious state problem in the Dirac equation \cite{zhang2022}.

Periodic potentials in quantum mechanics are frequently used to discuss atoms, ions, and molecules in a crystal, where the wave functions should satisfy Bloch's theorem \cite{kittel2018}. A simple example in textbooks is the periodic finite well potential, i.e., the Kronig-Penney model~\cite{kronig1931}, which can be semi-analytically solved and qualitatively describe the band structure of a crystal, including conducting, valence, and forbidden bands. Recently, the Kronig-Penney potential in the Schr\"{o}dinger equation was also solved numerically with the tail-cancellation method \cite{mishra2001}, infinite square well basis \cite{pavelich2015}, and the developed filter method \cite{abdurrouf2020}.

In this work, the FDM will be applied to solve the Kronig-Penney potential as a supplement in the lecture of quantum mechanics, with easily written matrix elements. The results will be compared to the semi-analytical solution and previous numerical methods. The influences of the strength and width of the Kronig-Penney potential on the eigenvalue and wave function will be investigated. Finally, as the extreme case of the Kronig-Penney potential, the Dirac comb potential \cite{cordoba1989} will be calculated. The structure of this paper is as follows. The analytical solution of the Kronig-Penney potential and the formula of FDM will be shown in Section \ref{sec2}. The numerical results of the Kronig-Penney potential using FDM will be presented in Section \ref{sec3}. Section \ref{sec4} will provide a summary and perspectives.

\section{Theoretical Framework}\label{sec2}
\subsection{Kronig-Penney Model}
\begin{figure}[h]
	\centering
	\includegraphics[width=0.6\textwidth]{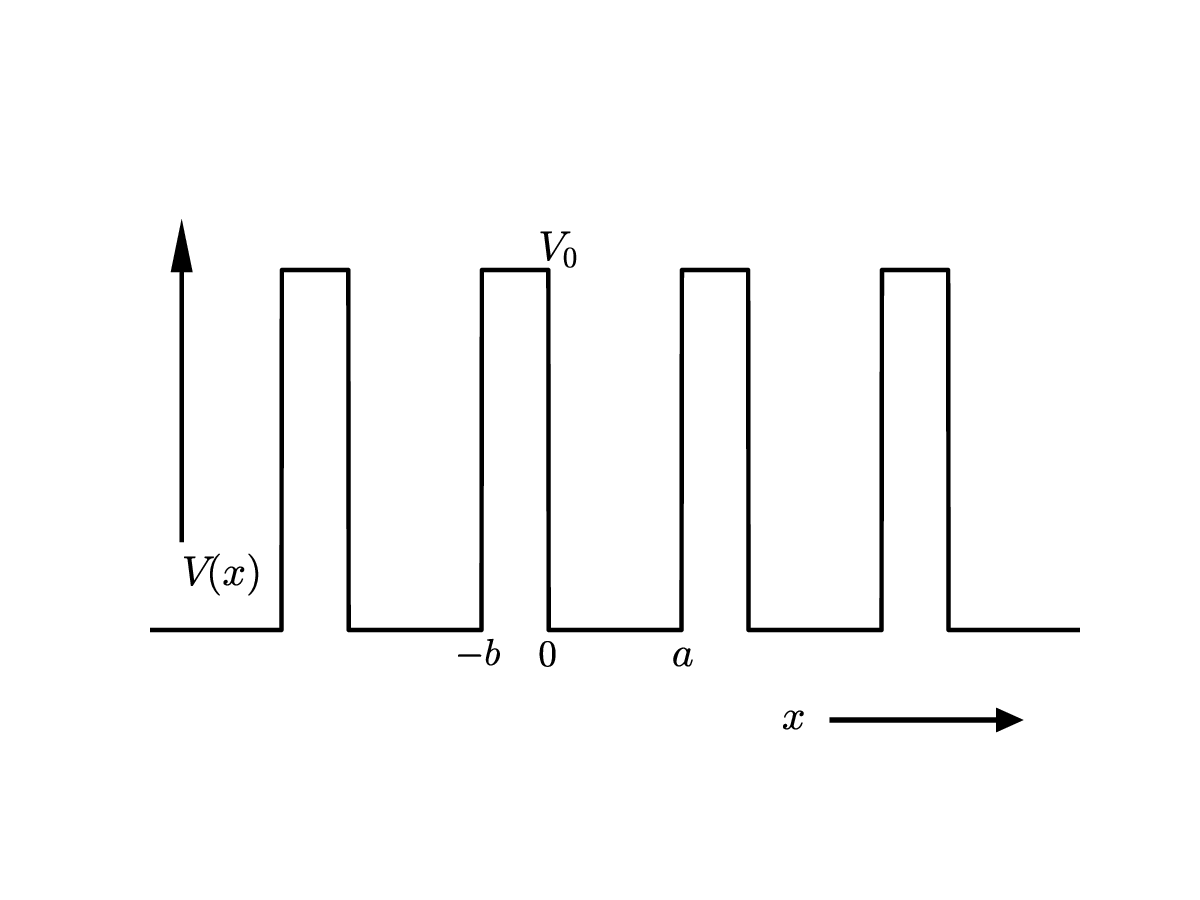}
	\caption{The KP Potential with height $V_0$ and periodic width $c=a+b$. }
	\label{Figure 1}
\end{figure}

The Kronig-Penney (KP) model is a periodic potential model first formulated in the 1930s by Kronig and Penney. It is represented by the one-dimensional potential shown in Fig. \ref{Figure 1}.
The KP potential in each period is expressed as
\begin{equation}
	V(x)=\begin{cases}
		V_0& -b\leq x\leq 0\\
		0&0< x\leq a.\\
	\end{cases}
\end{equation}
We define $c = a + b$ as the periodic length of potential. Since the potential is a periodic function of $x$ with a $c$, the Schr\"{o}dinger equation is invariant under space translations. According to Bloch's theorem, the wave function in a periodic potential, $\psi(x)$ and $\psi(x + c)$ must represent the same state. Therefore, $\psi(x + c)$ can differ from $\psi(x)$ only by a constant factor,
\begin{equation}
	\psi(x+c)=\beta\psi(x),
\end{equation}
where $\beta$ is a constant of magnitude unity,
\begin{equation}
	\beta=\exp\left(\frac{2\pi il}{n}\right),\qquad l=0,~1,~2,~3,~\dots,~n-1
\end{equation}
Through defining $\kappa={2\pi l}/{nc}$, we arrive at
\begin{equation}
		\psi(x+c)=e^{i\kappa c}\psi(x).
\end{equation}

Consider the case of a particle (with mass $m$ and total energy $E < V_0$) subject to the periodic potential described above. The Schr\"{o}dinger equation becomes,
\begin{equation}
	\begin{split}
		-\frac{\hbar^2}{2m}\frac{d^2}{dx^2}\psi(x)&=E\psi(x)~\quad\qquad0<x<a,\\
		-\frac{\hbar^2}{2m}\frac{d^2}{dx^2}\psi(x)+V_0\psi(x)&=E\psi(x)\qquad-b<x<0
	\end{split}
\end{equation}
and we define
\begin{equation}
	\begin{split}
	k^2_1&=\frac{2mE}{\hbar^2},\\
	k^2_2&=\frac{2m(V_0-E)}{\hbar^2}.
	\end{split}
	\label{Eq 9}
\end{equation}

In this work, we solve the Schr\"{o}dinger equation using Hartree atomic units with $e = m = \hbar = 1$ and $c = 137$. The solution for the wave function can be written as
\begin{equation}
	\begin{split}
		&\psi(x)=A\cos(k_1x)+B\sin(k_1x)~\qquad\qquad0< x<a,\\
		&\psi(x)=C\cosh(k_2x)+D\sinh(k_2x)\qquad-b< x<0.
	\end{split}
\end{equation}
The wave function and its first derivative must be continuous at the boundaries. At $x = 0$, we have
\begin{equation}
	\begin{split}
		A&=C,\\
		k_1B&=k_2D.
	\end{split}
\end{equation}
Furthermore, from Bloch's theorem, with $n=1$, we get the periodic boundary condition,
\begin{equation}
	\begin{split}
		\psi(a)&=e^{i\kappa c}\psi(-b),\\
		\psi'(a)&=e^{i\kappa c}\psi'(-b),
	\end{split}
\end{equation}
where $\kappa=2\pi l/c$, which leads to
\begin{equation}
	\begin{split}
		A\cos(k_1a)+B\sin(k_1a)&=e^{i\kappa c}[C\cosh(k_2b)-D\sinh(k_2b)],\\
		-k_1A\sin(k_1a)+k_1B\cos(k_1a)&=e^{i\kappa c}[-k_2C\sinh(k_2b)+k_2D\cosh(k_2b)].
	\end{split}
\end{equation}

These linear equations can be written as a matrix equation involving $A,~B,~C,~D$:
\begin{equation}
MX=0,
\end{equation}
where
\begin{equation}
	X=(ABCD)^T
\end{equation}
and
\begin{equation}
	M=\begin{pmatrix}
		1&0&-1&0\\
		0&k_1&0&-k_2\\
		\cos(k_1a)&\sin(k_1a)&-e^{i\kappa c}\cosh(k_2b)&e^{i\kappa c}\sinh(k_2b)\\
		-k_1\sin(k_1a)&k_1\cos(k_1a)&k_2e^{i\kappa c}\sinh(k_2b)&-k_2e^{i\kappa c}\cosh(k_2b)
	\end{pmatrix}.
\end{equation}

In order to obtain a nontrivial solution for $A,~B,~C,~D$, the determinant of above matrix, $M$ must be zero. It yields a transcendental equation for determining the energy eigenvalues,
\begin{equation}
	\frac{k_2^2-k_1^2}{2k_1k_2}\sinh(k_2b)\sin(k_1a)+\cosh(k_2b)\cos(k_1a)=\cos[\kappa c],
\end{equation}
which can be written for the bound state as
\begin{equation}
	\frac{V_0-2E}{2\sqrt{E}\sqrt{V_0-E}}\sinh(\sqrt{2(V_0-E)}b)\sin(\sqrt{2E}a)+\cosh(\sqrt{2(V_0-E)}b)\cos(\sqrt{2E}a)=\cos[\kappa c].
	\label{Eq 19}
\end{equation}
When $E>V_0$, using the identities $\sinh iz=i\sin z$ and $\cosh iz=\cos z$ \cite{mcquarrie1996}, Eq. (\ref{Eq 19}) will become as
\begin{equation}
	\frac{V_0-2E}{2\sqrt{E}\sqrt{E-V_0}}\sin(\sqrt{2(E-V_0)}b)\sin(\sqrt{2E}a)+\cos(\sqrt{2(E-V_0)}b)\cos(\sqrt{2E}a)=\cos[\kappa c].
	\label{Eq 20}
\end{equation}

\begin{figure}[h]
	\centering
	\includegraphics[width=0.6\textwidth]{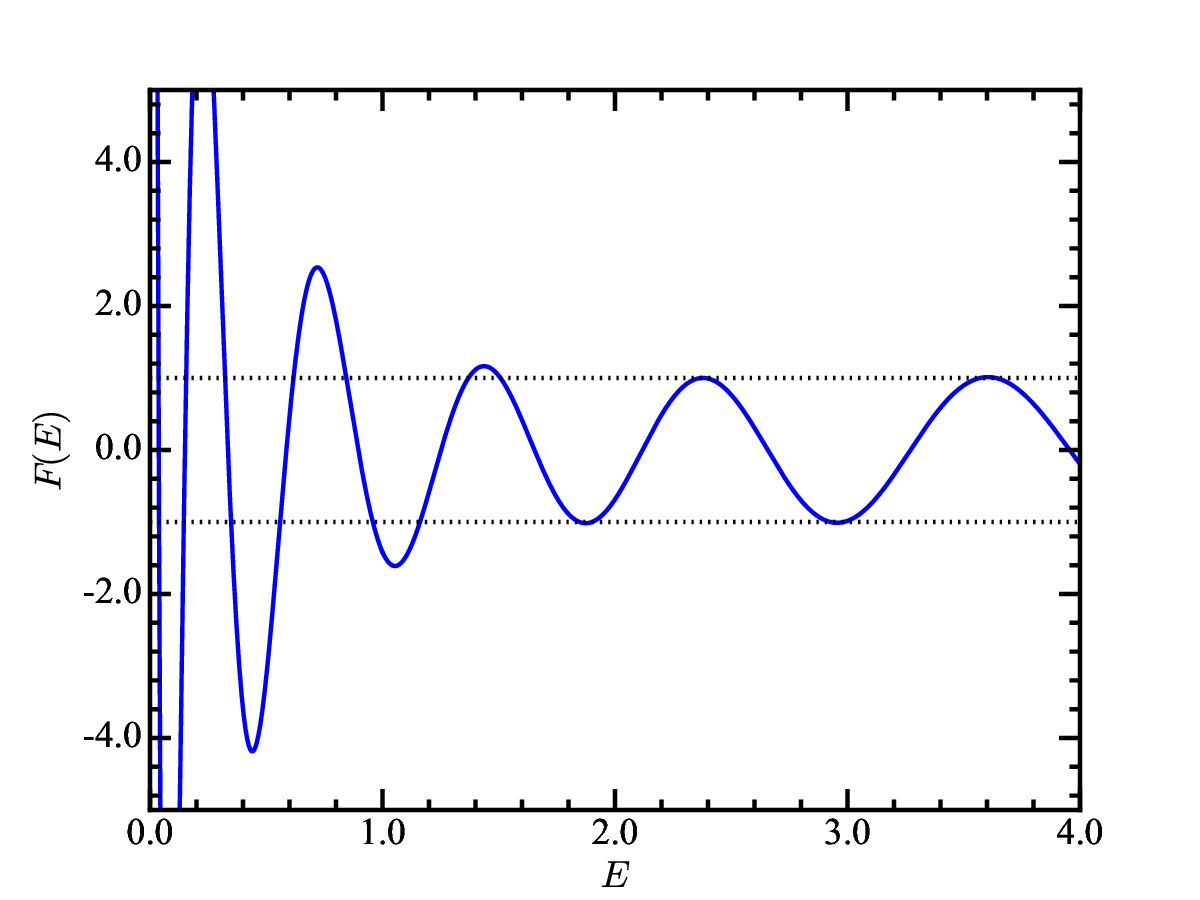}
	\caption{The $F(E)$ as a function of $E$. The dashed lines corresponds to the boundary $|\cos(\kappa c)|=1$.}
	\label{Figure2}
\end{figure}

Then, Eqs. (\ref{Eq 19}) and (\ref{Eq 20}) can be summarized as
\begin{equation}
	F(E)=\cos[\kappa c].
	\label{Eq 21}
\end{equation}
This equation provides the eigenvalue $E$ as a function of $\kappa c$. Eq. (\ref{Eq 21}) can be easily solved using a graphical method. We plot $F(E)$ as a function of $E$ with the potential parameters $V_0 = 0.5$, $a = 10$, and $b = 2$ as an example in Fig. \ref{Figure2}.
The dashed lines denote $\cos(\kappa c) = \pm 1$, respectively. The existence of an eigenvalue $E$ requires that the magnitude of $F(E)$ be between $-1$ and $+1$. If $|F(E)| > 1$, there is no bound state, which corresponds to the forbidden bands.

\subsection{The Finite Difference Method}
The finite difference method (FDM) is a powerful numerical approach for solving differential equations. The mathematical definition of a derivative is expressed in terms of a limit,
\begin{equation}
	\frac{df(x)}{dx}=\lim_{\Delta x\to 0}\frac{f(x+\Delta x)-f(x)}{\Delta x}.
\end{equation}
Thus, we can approximately describe the above derivative by using the difference between the wave functions on two neighboring lattice points with a small finite step $h$,
\begin{equation}
	\frac{df(x)}{dx}\approx\frac{f(x+h)-f(x)}{h}.
\end{equation}
The higher derivatives can be described in a similar form with the help of Taylor expansion. Therefore, the second derivative is given by the function values at three neighboring points $x-h,~x, ~x+h$,
\begin{equation}
	\frac{ d^2f(x)}{ dx^2}\approx\frac{f(x-h)-2f(x)+f(x+h)}{h^2},
\end{equation}
which is also named as central difference formula. Therefore, the time independent Schr\"{o}dinger equation in one dimension can be discretized in a coordinate box with a fixed step of size $h$. The wave function can be written as  $\psi(0),~\psi(h),~\psi(2h),~\cdots,~\psi((N+1)h)$. On each lattice point, the Schr\"{o}dinger equation is shown as
\begin{equation}
	-\frac{\hbar^2}{2m}\frac{\psi(i-h)-2\psi(i)+\psi(i+h)}{h^2}+V(i)\psi(i)=E\psi(i),\qquad i=h,~2h~\cdots~ Nh.
\end{equation}

For a finite-range potential, the wave function at the boundaries of a box should be zero. Therefore, $\psi(0)$ and $\psi((N+1)h)$ should be removed. The Schr\"{o}dinger equation is then transformed into a matrix form,
\begin{equation}
	\left[-\frac{\hbar^2}{2mh^2}\left(\begin{matrix}
		-2&1&&&\\
		1&-2&1&&\\
		&1&-2&\ddots&\\
		&&\ddots&\ddots&1\\
		&&&1&-2\\
	\end{matrix}\right)+\left(\begin{matrix}
	&V(h)&&&\\
	&&V(2h)&&\\
	&&&V(3h)&\\
	&&&\ddots&\\
	&&&&V(Nh)\\
\end{matrix}\right)\right]\left(\begin{matrix}
	\psi(h)\\\psi(2h)\\\psi(3h)\\\vdots\\\psi(Nh)
	\end{matrix}\right)=E\left(\begin{matrix}
	\psi(h)\\\psi(2h)\\\psi(3h)\\\vdots\\\psi(Nh)
	\end{matrix}\right).
\end{equation}
Thus, the left side of the equation can be written as an $n$-order sparse matrix, to be diagonalized to provide the eigenvalues  as the eigen energies, and the eigen vectors are the corresponding wave functions. By using FDM, we have transformed the eigenvalue problem of the Schr\"{o}dinger equation into a matrix equation, which can be easily solved using techniques from linear algebra.

When a periodic potential is considered, the wave function at the boundary points must satisfy Bloch's theorem, i.e., $\psi(0) = e^{-ikc}\psi(Nh)$ and $\psi((N+1)h) = e^{ikc}\psi(h)$ and the matrix equation must be adjusted accordingly. Thus, the second difference terms at the boundary points are transformed into,
\begin{equation}
	\begin{split}
		&\frac{ d^2}{ dx^2}\psi(x)|_{x=h}=\frac{e^{-ikc}\psi(Nh)-2\psi(h)+\psi(2h)}{h^2},\\
		&\frac{d^2}{ dx^2}\psi(x)|_{x=Nh}=\frac{\psi((N-1)h)-2\psi(Nh)+e^{i\kappa c}\psi(h)}{h^2},
	\end{split}
\end{equation}
where $\kappa=2\pi l/nc$.\\

Therefore, the Schr\"{o}dinger equation with a periodic potential in FDM is written as
\begin{equation}
	\left[-\frac{\hbar^2}{2mh^2}\left(\begin{matrix}
		-2&1&&&e^{-i\kappa c}\\
		1&-2&1&&\\
		&1&-2&\ddots&\\
		&&\ddots&\ddots&1\\
		e^{i\kappa c}&&&1&-2\\
	\end{matrix}\right)+\left(\begin{matrix}
	&V(h)&&&\\
	&&V(2h)&&\\
	&&&V(3h)&\\
	&&&\ddots&\\
	&&&&V(Nh)\\
\end{matrix}\right)\right]\left(\begin{matrix}
		\psi(h)\\\psi(2h)\\\psi(3h)\\\vdots\\\psi(Nh)
	\end{matrix}\right)=E\left(\begin{matrix}
		\psi(h)\\\psi(2h)\\\psi(3h)\\\vdots\\\psi(Nh)
	\end{matrix}\right),
\end{equation}
which can be also diagonalized to obtain the eigen energies and the corresponding wave functions.

\section{Numerical Result and Discussion}\label{sec3}
\subsection{The Kronig-Penney Potential}
\begin{figure}[h]
	\centering
	\includegraphics[width=0.6\textwidth]{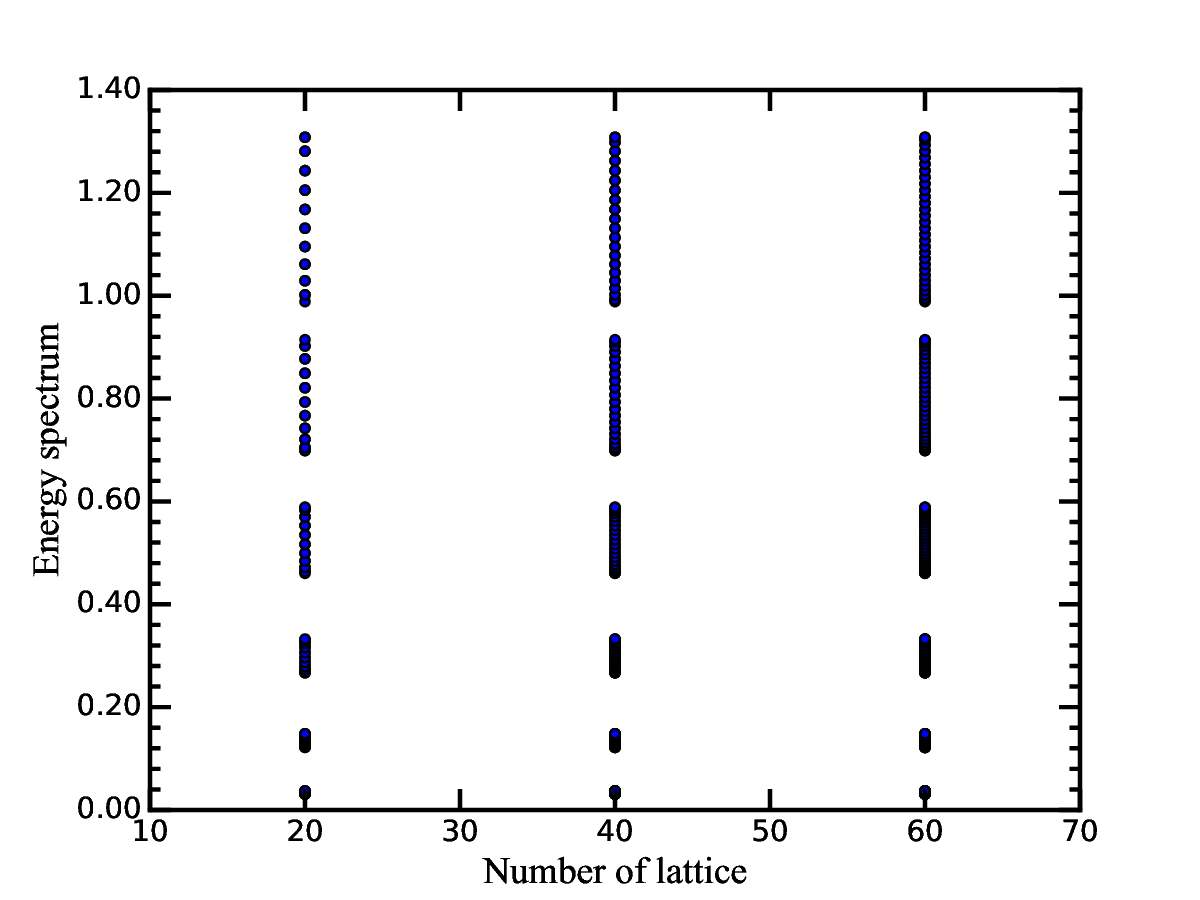}
	\caption{The energy spectrum with different numbers of periodic lattice.}
	\label{Figure3}
\end{figure}

First, a KP potential with $a = 2$, $b = 10$, and $V_0 = 0.5$ is solved using the FDM framework as an example, where the box is uniformly separated into $N=10000$ points and  step size is $h=12/10000$. The eigenvalues and wave functions are strongly dependent on the number of periods, $n$. We calculate the eigenvalues of the first six energy bands for $n=20,~40$, and $60$, respectively as shown in Fig. \ref{Figure3}. As the number of lattice points increases, the density of states becomes larger. When $n = 60$, the band structure is clearly displayed.

\begin{figure}[H]
	\centering
	\includegraphics[width=0.6\textwidth]{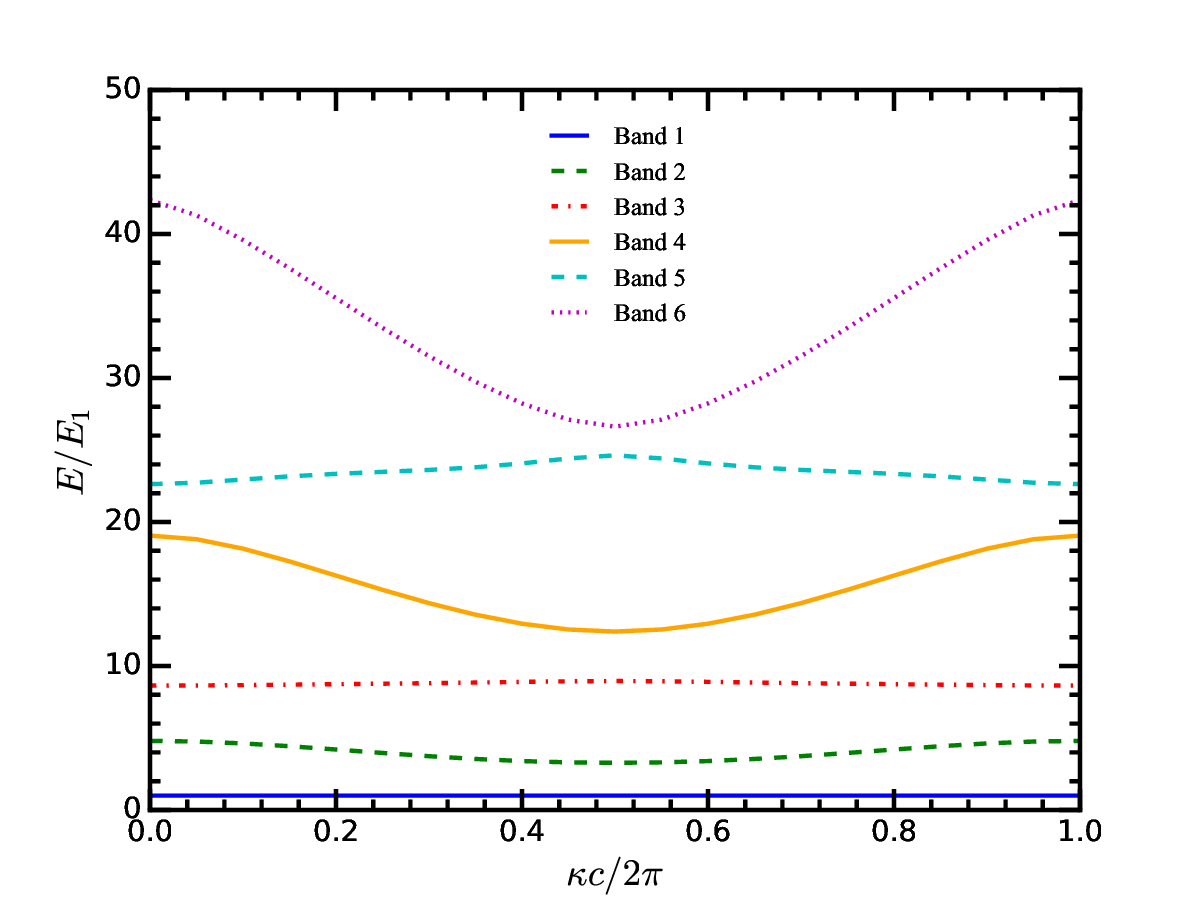}
	\caption{The band structures for KP potential as a function of wave vector.}
	\label{Figure4}
\end{figure}

In a realistic crystal, the number of lattice points can be considered as an infinite large number. The wave vector $\kappa$ is continuous and satisfies $0 \leq \kappa c / 2\pi \leq 1$. The energy spectrum is expressed as a function of $\kappa c / 2\pi$, as shown in Fig.~\ref{Figure4}. The six lowest bands are plotted and normalized with respect to the ground state energy, $E_1$.

It can be observed that the odd-numbered energy bands exhibit a convex function distribution, with the maximum value at $\kappa c / 2\pi = 0.5$. In contrast, the even-numbered energy bands exhibit a concave function distribution, with the minimum value in the middle region. In bands $4$ and $6$, the eigenvalue changes significantly within the same band.

On the other hand, the eigenvalue solutions for the KP potential can be semi-analytically obtained by solving the transcendental equations in Eqs. (\ref{Eq 19}) and (\ref{Eq 20}). These solutions are compared with those obtained by using the filter method \cite{abdurrouf2020}, which is based on the idea that any wave packet within a defined space can be expressed as a superposition of all wave functions, and with those obtained by using FDM, as shown in Table \ref{Table1}.

\begin{table}[htbp]
	\centering
	\caption{The eigenvalues of KP potential solved by different methods.}
	\setlength{\tabcolsep}{4mm}{
		\begin{tabular}{cccc}
			\hline \hline  
			Energy band&Analytic results&Filter method\cite{abdurrouf2020}&FDM \\ 
			\hline
			1&0.030796 - 0.037141&0.030007 - 0.036291&0.030804 - 0.037141\\
			2&0.121302 - 0.148295&0.120179 - 0.146878&0.121334 - 0.148290\\
			3&0.266337 - 0.332651&0.260877 - 0.329653&0.266391 - 0.332637\\
			4&0.459674 - 0.588875&0.453034 - 0.584022&0.459723 - 0.588843\\
			5&0.698950 - 0.915166&0.693768 - 0.908540&0.698941 - 0.915100\\
			6&0.989655 - 1.309390&0.988431 - 1.301516&0.989543 - 1.309269\\
			7&1.342598 - 1.767130&1.347481 - 1.761236&1.342370 - 1.766798\\
			\hline
	\end{tabular}}
	\label{Table1}
\end{table}
It can be observed that the FDM solution has higher accuracy than that from the filter method and is closer to the analytical result. In principle, accuracy can be improved by increasing the discretized step size, but this also increases the computation time. We have checked the convergence of the results with respect to the discretized step size $N$. The solution of the diagonalizing matrix becomes stable for $N = 10000$ and $h=12/10000$.

\begin{figure}[ht]
	\centering
	\includegraphics[width=0.6\textwidth]{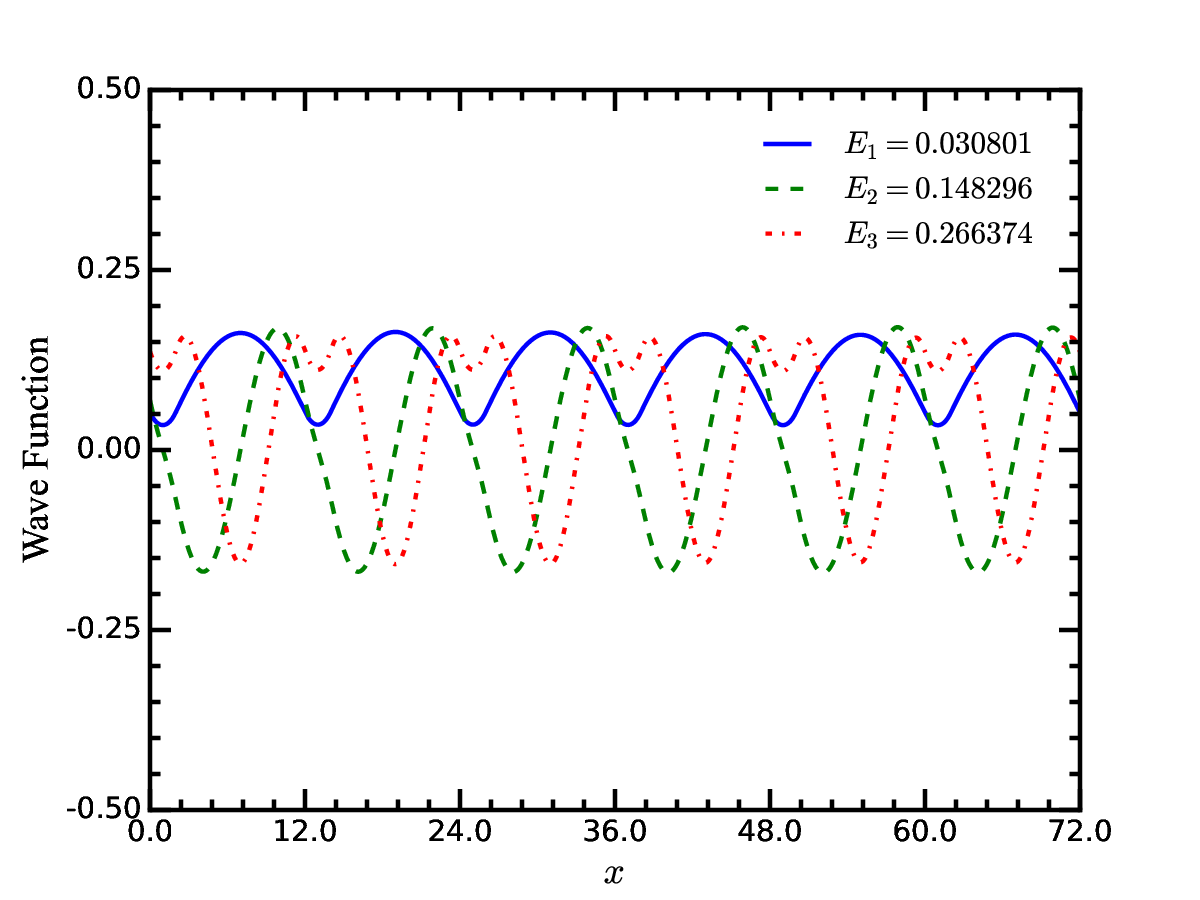}
	\caption{The wave functions of lowest three states for KP potential with  $\kappa c/2\pi=1$.}
	\label{Figure5}
\end{figure}

We also consider a box including six periodic potentials, i.e., $R = 72$. The wave functions of the lowest three states for $\kappa c / 2\pi = 1$ are plotted in Fig. \ref{Figure5}. The periodicity in the wave function is clearly observed in the lattice space, demonstrating that the FDM is effective for periodic potentials in the Schr\"{o}dinger equation.

\subsection{The influences of height and width of KP potentials on energy structure}
\begin{figure}[ht]
	\centering
	\includegraphics[width=0.8\textwidth]{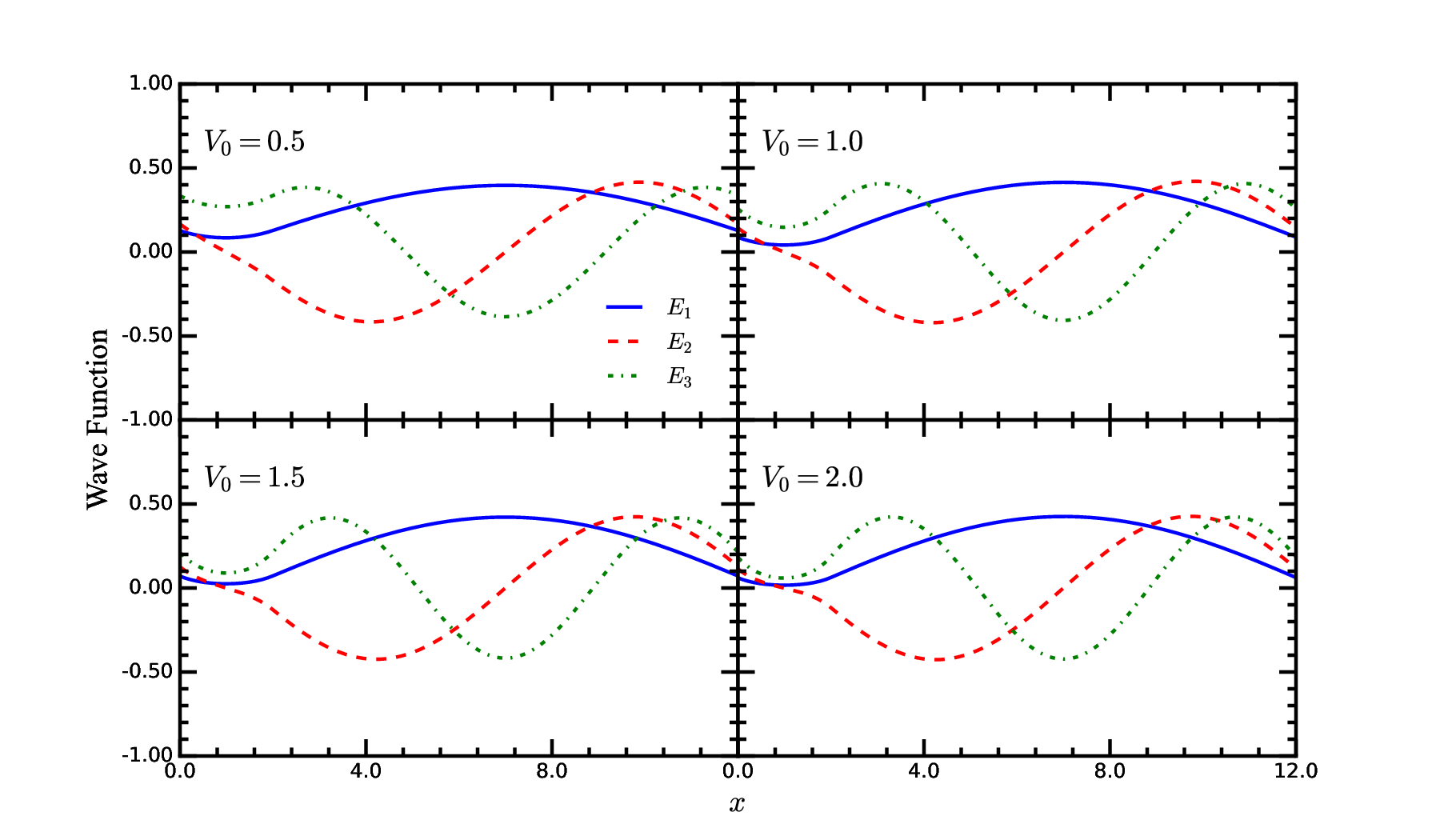}
	\caption{The wave functions for different heights of KP potential.}
	\label{Figure6}
\end{figure}

We have successfully solved the Schr\"{o}dinger equation for the KP potential using the FDM. Next, we will further investigate the effects of the height and width of the KP potential on the wave functions and eigenvalues. The potential heights $V_0$ are varied from $0.5$ to $1.0$, $1.5$, and $2.0$. The corresponding wave functions for $\kappa c / 2\pi = 1$ are shown in Fig. \ref{Figure6}. Generally, the shapes of these wave functions with different potential heights are quite similar. As the height increases to be infinite, the wave functions approach the exact sine function, which represents the wave function of an infinite well potential.

\begin{figure}[H]
	\centering
	\includegraphics[width=0.8\textwidth]{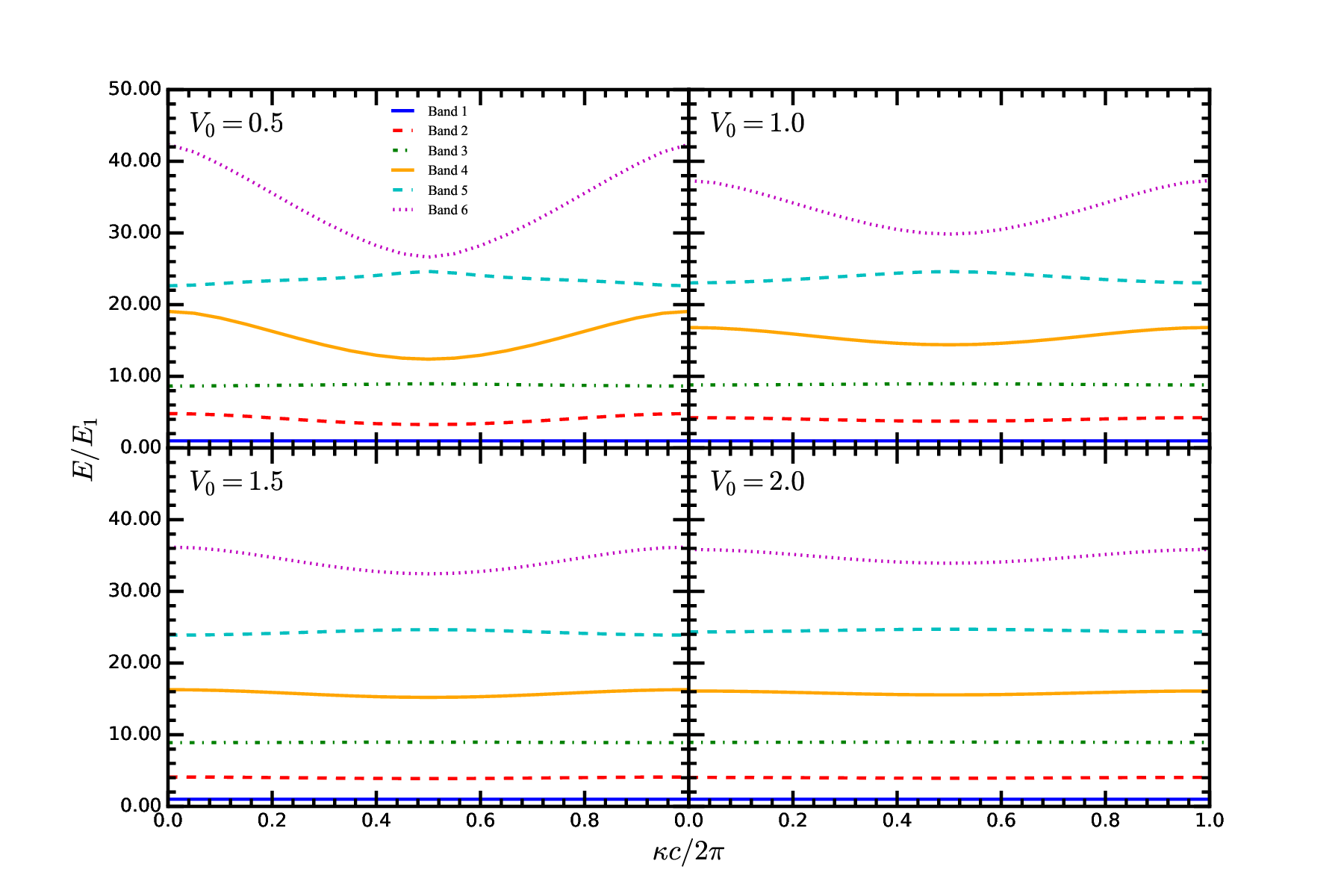}
	\caption{The band structures with different potential heights in the KP potential.}
	\label{Figure 7}
\end{figure}

Correspondingly, we can also calculate the band structure of a single particle in the periodic potential to more easily observe the influence of the potential height. The results are shown in Fig. \ref{Figure 7}. If the height of the potential is lower, the eigenvalues in each band change significantly with the wave vectors, $\kappa c / 2\pi$, especially for the higher energy levels. For $V_0 = 0.5$, the energy difference between the two boundary points and the central point is $15E_1$ at band 6. As the potential height increases, the eigenvalues become flatter and do not change much with the wave vectors. In fact, they start to resemble the energy levels of an infinite well potential, $n^2 \pi^2 / 2$.

\begin{figure}[h]
	\centering
	\includegraphics[width=0.8\textwidth]{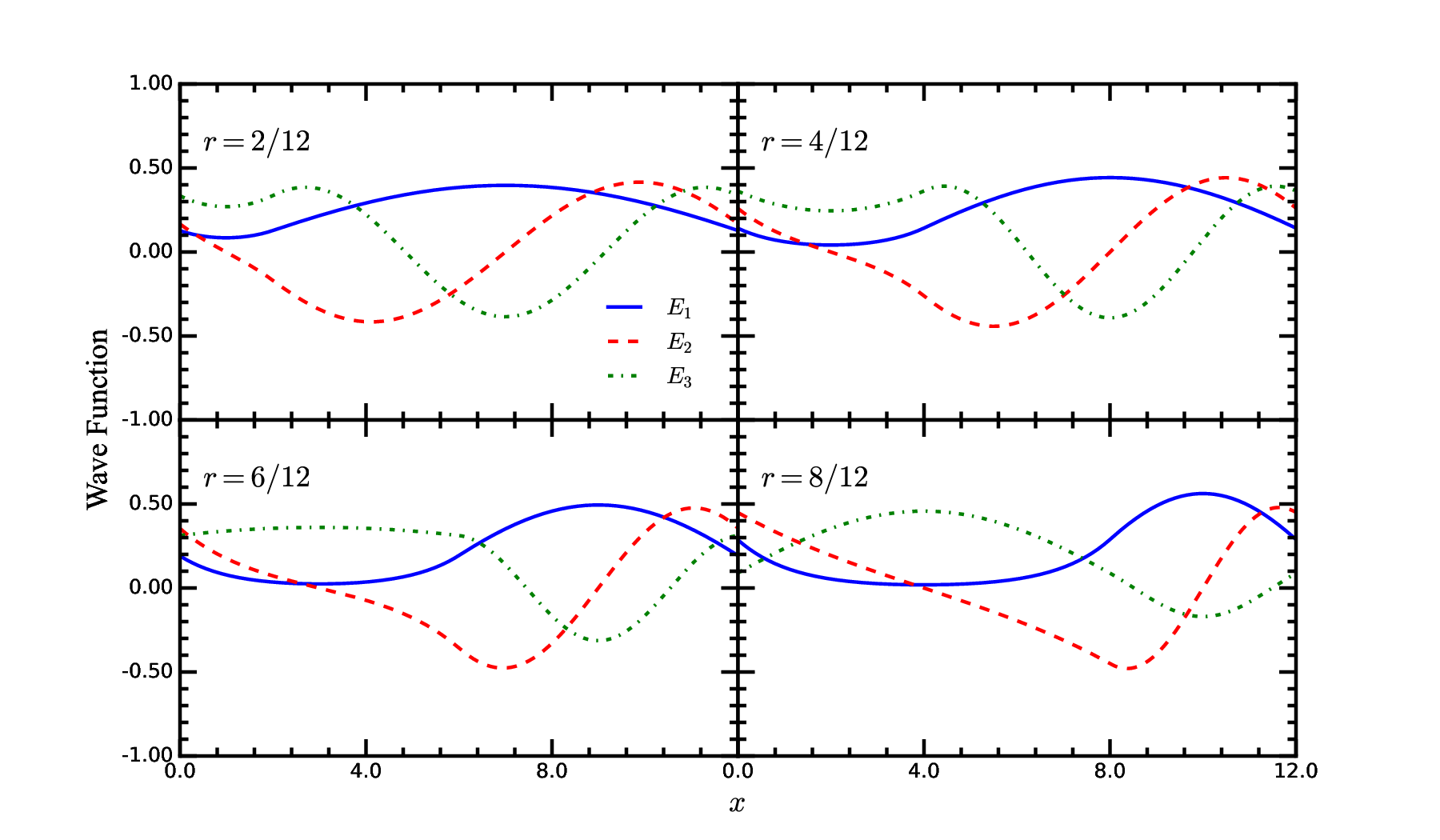}
	\caption{The wave functions for different potential width ratios in KP potential.}
	\label{Figure 8}
\end{figure}

\begin{figure}[H]
	\centering
	\includegraphics[width=0.8\textwidth]{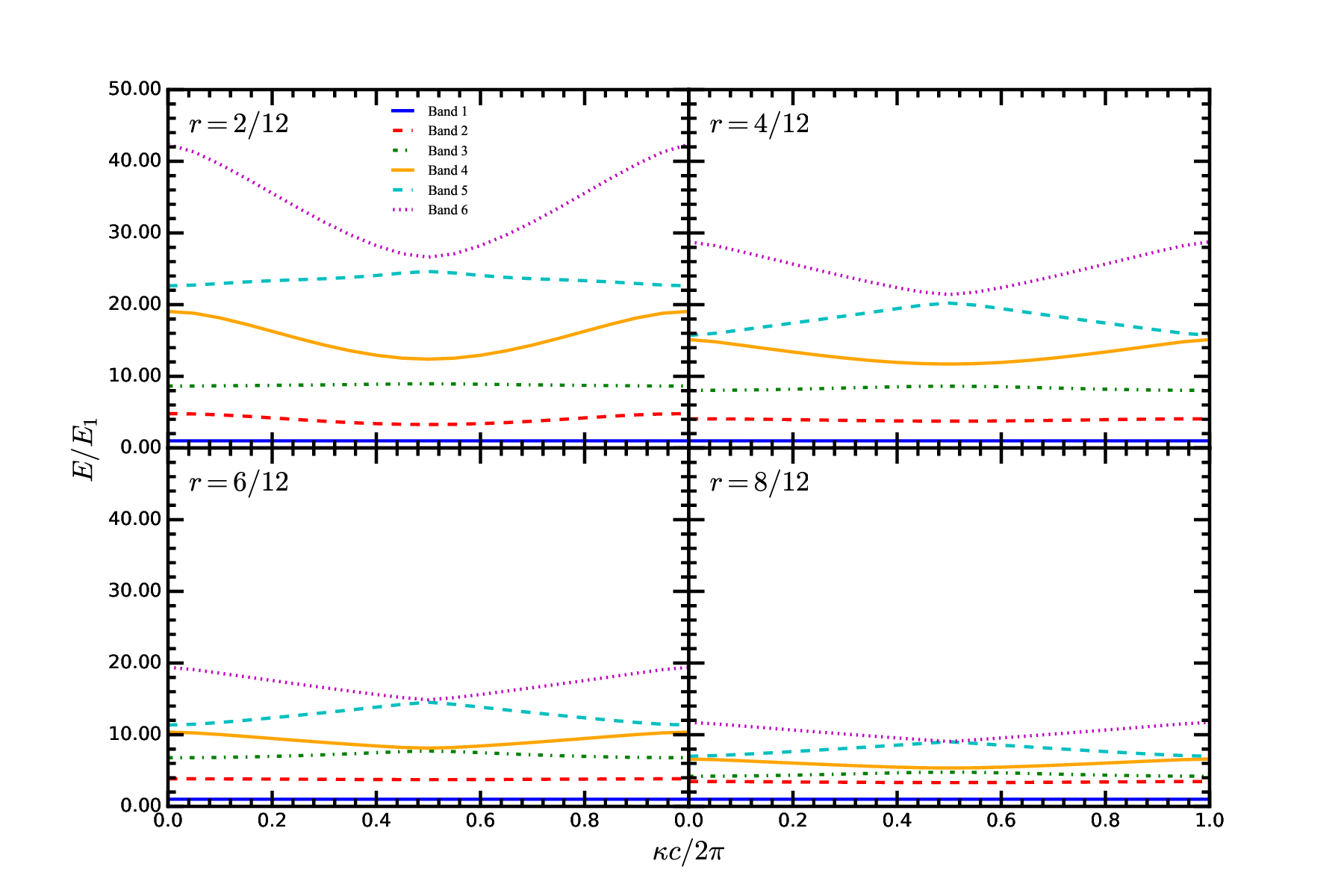}
	\caption{The energy band structures for different potential width ratios in KP potential.}
	\label{Figure 9}
\end{figure}

Similarly, we also discuss the influence of the potential width on the wave functions and energy structures, where we set $r = a/c = 2/12$, $4/12$, $6/12$, and $8/12$, respectively. The wave functions and eigenvalues are plotted in Figs. \ref{Figure 8} and \ref{Figure 9}, respectively. The wave functions undergo significant changes with the increment of potential width. The wave function becomes flatter in the potential interaction region. For the case $a/c = 8/12$, the probability of finding the particle in the ground state is almost constant. Furthermore, the eigenvalues of each band decrease as the potential width ratio increases, and different energy bands get closer. This is similar to the fact that the eigenvalues decrease with increasing width in the infinite well potential. We also notice that the gaps between band $3$ and band $4$,  gaps between band $5$ and band $6$, almost disappear at $\kappa c = \pi$, becoming continuous, which is similar to the Dirac cone structure.


\subsection{The Dirac Comb Potential}
\begin{figure}[h]
	\centering
	\includegraphics[width=0.6\textwidth]{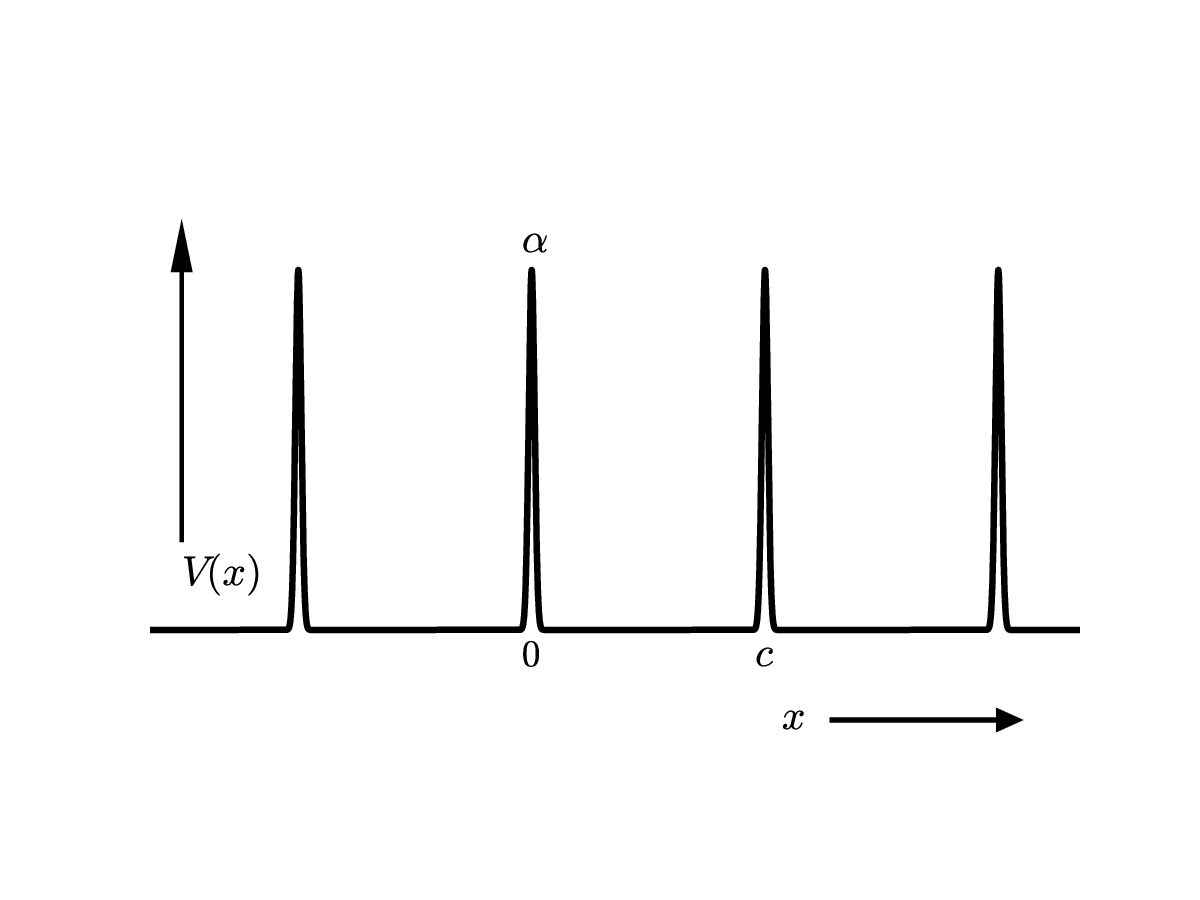}
	\caption{The Dirac Comb Potential as a function of distance.}
	\label{Figure 10}
\end{figure}

If the height and width in the KP potential have the limit value $V_0\rightarrow \infty$ and $b\rightarrow 0$, it will become a new potential named as Dirac comb potential, which is defined as,
\begin{equation}
	V(x)=\alpha\sum\limits_{j=0}\limits^{N-1}\delta(x-jc),
\end{equation}
where $\alpha$ is $\delta$ potential strength and $c$ is the lattice distance. The shape of one-dimension Delta comb potential is shown in Fig. \ref{Figure 10}.
\\

The analytical solution of the Dirac comb potential can be obtained through a process similar to that used for the KP potential. The only difference is that $b$ is now $0$. Therefore, Eq. (\ref{Eq 19}) will be evaluated as
\begin{equation}\label{dcpe}
	\alpha\frac{\sin(\sqrt{2E}c)}{\sqrt{2E}}+\cos(\sqrt{2E}c)=\cos[\kappa c].
\end{equation}
The eigenvalue, $E$ can be solved with the graphically method  in this equation when the wave vector $\kappa c$ is fixed.

In the present FDM framework, the Dirac comb potential can be numerically solved using a similar scheme to the KP potential in the Schr\"{o}dinger equation. The only problem is that the $\delta$ function cannot be directly implemented in the program. Therefore, we assume that $V_0$ in the KP potential is very large while $b$ is small enough. The limit of $b$ in FDM is $c/N$, where $N$ is the total number of lattice points. When we consider the strength of the Dirac comb potential as $\alpha=1$, it corresponds $V_0b=1$ in the KP potential from Eq. (\ref{Eq 19}). Under this condition, the wave functions and eigenvalues of the Dirac comb potential are solved within the FDM framework.

The comparison between the analytical results from Eq. (\ref{dcpe}) and the numerical solutions for the eigenvalues of the Dirac comb potential is presented in Table \ref{Table 4}. There are also energy bands in the Dirac comb potential, and the eigenvalues from FDM are consistent with the calculations from the analytical formula.

\begin{table}[htbp]
	\centering
	\caption{The eigenvalues at different energy bands solved by analytical method and FDM.}
	\setlength{\tabcolsep}{2mm}
	\begin{tabular}{ccc}
		\hline \hline  
		Energy band&Analytic results&FDM \\ 
		\hline 
		1&0.025296 - 0.034269&0.025296 - 0.034263\\
		2&0.102488 - 0.137078&0.102488 - 0.137050\\
		3&0.234797 - 0.308425&0.234789 - 0.308363\\
		4&0.425939 - 0.548311&0.425908 - 0.548202\\
		5&0.679069 - 0.856736&0.679005 - 0.856565\\
		6&0.996470 - 1.233701&0.996355 - 1.233454\\
		\hline \hline  
	\end{tabular}
	\label{Table 4}
\end{table}

\begin{figure}[h]
	\centering
	\includegraphics[width=0.6\textwidth]{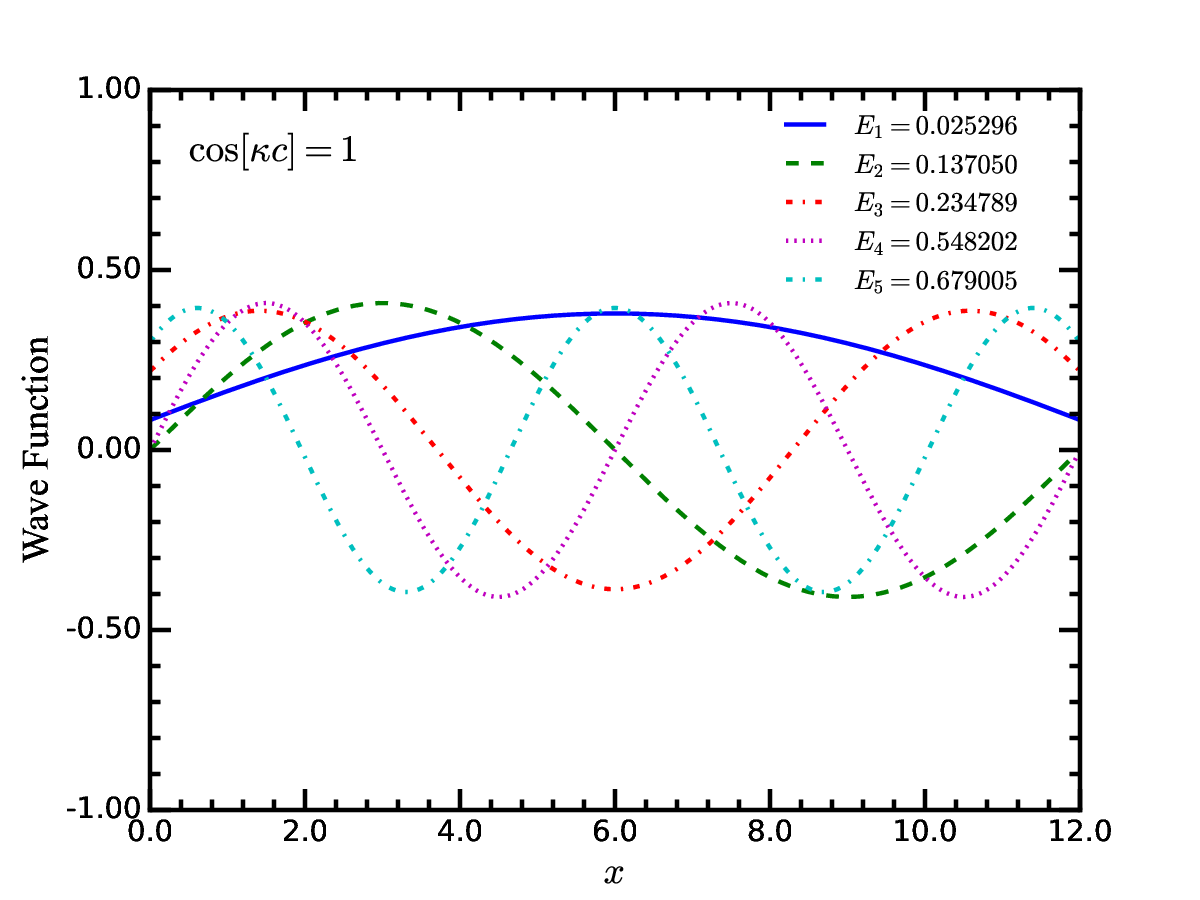}
	\caption{The wave functions for Dirac comb potential from FDM.}
	\label{Figure 11}
\end{figure}

Finally, the wave functions of the Dirac comb potential are shown in Fig. \ref{Figure 11} for the case where $\cos[\kappa c] = 1$. These functions resemble sine functions, similar to the solution of an infinite well potential. 
\\

\section{Conclusion}\label{sec4}
The KP potential, a classical periodic potential in quantum mechanics textbooks, is numerically solved using the finite difference method (FDM) in the Schr\"{o}dinger equation by considering periodic boundary conditions. The eigenvalues at different bands are comparable with the analytical results. This demonstrates that FDM is an efficient, simpler, and high-precision numerical method for solving periodic potentials compared to other numerical techniques.

We also discussed the influences of the height and width of the KP potential on the wave functions and eigenvalues within the present framework. It was found that the energy eigenvalues at each band change significantly with the wave vector when the potential height is small, while they become flatter for higher heights and resemble the case of an infinite well potential. Furthermore, if the width of the potential increases, the energy gap between different bands greatly reduces. Between some bands, the energy gap even disappears.

Finally, as the limit case of KP potential, the Dirac comb potential is also considered, when its height is  infinite and the width is extremely narrow. The FDM can also numerically provide exact solutions compared to the analytical results. This proves that the FDM is highly effective, and the present work offers a valuable supplement to traditional teaching materials on the KP and Dirac comb potentials. In practical applications, many physical systems should extend beyond one dimension and the Schr\"{o}dinger equation. Future work will investigate $2$- and $3$-dimensional periodic potentials and the relativistic Dirac equation.

\section{Acknowledgments}
This work was supported in part by the National Natural Science Foundation of China (Grant  Nos. 11775119 and 12175109), the Natural Science Foundation of Tianjin (Grant  No: 19JCYBJC30800) , and the Natural Science Foundation of Guangdong Province (Grant  No: 2024A1515010911). 

\bibliographystyle{apsrev4-2}
\bibliography{ref1.bib}
\end{document}